\documentclass[12pt]{article}
\usepackage{amsmath,amssymb}
\usepackage{graphicx}
\usepackage[format=hang,labelfont=bf,textfont=small,singlelinecheck=false,justification=raggedright,margin={12pt,12pt},figurename=Fig.]{caption}
\usepackage{titlesec}
\usepackage{textcomp}
\usepackage{dcolumn}
\usepackage{bm}
\usepackage[version=3]{mhchem}
\usepackage{setspace}

\newcommand{\tc}{\ensuremath{T_\mathrm{c}}}

\newcommand{\etal}{\textit{et al.}}
\newcommand{\simm}{$\sim$}

\newcommand{\cccm}{\textit{Cccm}} 
\newcommand{\rmm}{\textit{R}3\textit{m}} 
\newcommand{\imm}{\textit{Im}$\overline{3}$\textit{m}} 

\usepackage[top=30truemm,bottom=30truemm,left=25truemm,right=25truemm]{geometry}

\makeatletter
\def\affiliation#1{\gdef\@affiliation{#1}}
\def\abstract#1{\gdef\@abstract{#1}}
\def\graphabst#1{\gdef\@graphabst{#1}}
\def\keywords#1{\gdef\@keywords{#1}}
\def\corresp#1{\gdef\@corresp{#1}}

\newcommand{\MakeTitle}{
  \newpage
  \null
  \vskip 2em%
  \begin{center}%
  \Large \@title\par
  \vskip 1em%
  \large \@author
  \end{center}
  \noindent\@affiliation\par
  \vskip 1em%
  \noindent\@corresp\par
  \vskip 1em%
  \noindent\@abstract\par
  \noindent\@graphabst\par
  \vskip 1em%
  \noindent\@keywords\par
}
\makeatother

\newcommand*{\TitleFont}{%
      \usefont{\encodingdefault}{\rmdefault}{}{n}%
      \fontsize{18}{12}%
      \selectfont}

\titleformat{\section}
  {\normalfont\fontsize{10}{11}\bfseries}{\thesection.}{2pt}{}
  \titlespacing*{\section}{0pt}{12pt}{6pt}

\titleformat{\subsection}
  {\normalfont\fontsize{10}{10}\bfseries}{\thesubsection.}{2pt}{}
  \titlespacing*{\subsection}{0pt}{6pt}{0pt}
  
\titleformat{\subsubsection}
  {\normalfont\fontsize{10}{10}\bfseries}{\thesubsubsection.}{2pt}{}
  \titlespacing*{\subsubsection}{0pt}{6pt}{0pt}
  
\normalsize

\title{\TitleFont Superconductivity of Pure \ce{H3S} Synthesized from Elemental Sulfur and Hydrogen}
\author{Harushige Nakao$^{\, 1,\ast }$, Mari Einaga$^{\, 1}$, Masafumi Sakata$^{\, 1}$, Masaomi Kitagaki$^{\, 1}$, Katsuya Shimizu$^{\, 1}$, Saori Kawaguchi$^{\, 2}$, Naohisa Hirao$^{\, 2}$, and Yasuo Ohishi$^{\, 2}$}
\affiliation{$^{1}$KYOKUGEN, Graduate School of Engineering Science, Osaka University, Toyonaka, Osaka 560-8531, Japan \\
$^{2}$JASRI, Sayo, Hyogo 679-5198, Japan}
\corresp{$^{\ast }$nakao@hpr.stec.es.osaka-u.ac.jp}

\abstract{\textbf{Abstract}: Superconductive \ce{H3S} synthesized from \ce{H2S} is very poorly crystallized, and has excess sulfur as impurity (3\ce{H2S} $\rightarrow$ 2\ce{H3S} + \ce{S}). The phase transition process undergoes in sulfur excess condition, which might cause hydrogen deficiency. The influence of hydrogen deficiency is not clear. Therefore investigation on the superconductivity in \ce{H3S} with no hydrogen deficiency is demanded. Two groups performed synthesis of \ce{H3S} from elemental sulfur and hydrogen (direct synthesis) and their results have shown that no hydrogen deficiency is caused when the direct synthesis is performed under hydrogen excessive condition. However, no measurements of superconductivity has been carried out because of the major technical difficulties in hydrogen experiments in diamond anvil cells (DACs). Here, we report the first electrical resistance measurements in superconductive \ce{H3S} synthesized from elemental sulfur and hydrogen (3\ce{H2} + 2\ce{S} $\rightarrow$ 2\ce{H3S}). Our powder X-ray diffraction (PXRD) using a synchrotron X-ray revealed that synthesized \ce{H3S} has much improved crystalline quality and no sulfur as reported in previous works. We observed a superconducting transition with a sharp drop of the resistance at $T_{onset}$ = 200 K and we obtained the highest $T_{offset}$ of 186 K in S-H system.}

\graphabst{
\begin{center}
\end{center}
}

\begin{document}

\onecolumn
\MakeTitle

\onecolumn

High temperature superconductivity near 200 K has been recently discovered in compressed hydrogen sulfide (\ce{H2S}) by Drozdov \etal\ \cite{Drozdov2015}.
The maximum superconducting transition temperature (\tc) observed was 203 K at 155 GPa, and it broke the record of \tc\ which was previously held by cuprate superconductor \cite{Schilling1983}.
This experimental discovery substantiated possibility of high-temperature superconductivity in hydrogen-dominant materials under high pressures proposed by Ashcroft \cite{Ashcroft2004} and has shown us a new route to the possible room temperature (RT) superconductivity.

\begin{figure}[hbt]
\includegraphics[width=8.6cm]{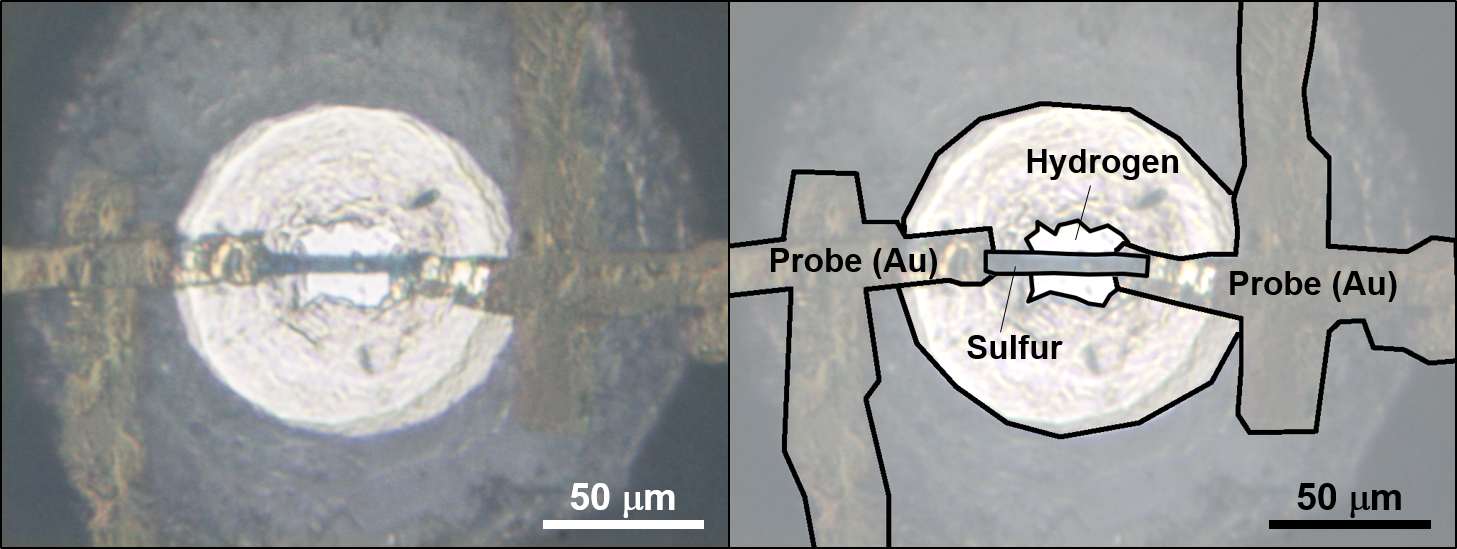}
\centering
\caption{A photo image of the sample after loading at 25 GPa (RT). Gold (Au) probes are connected to a rectangle-shaped sulfur and hydrogen is beneath the sulfur. The photo was taken in transmission and reflection illumination.}
\label{S+H007_AF_loading}
\end{figure}

\begin{figure}[hbt]
\includegraphics[width=8.6cm]{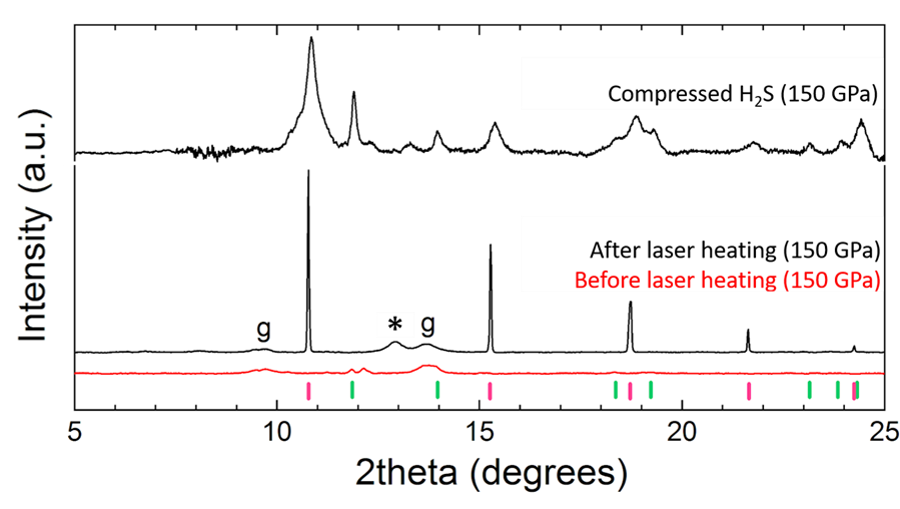}
\centering
\caption{Integrated PXRD patterns obtained with subtraction of the background for compressed \ce{H2S} \cite{Einaga2016} (top), \ce{H3S} synthesized from sulfur and \ce{H2} (bottom, black solid line) and S+\ce{H2} mixture before laser heating (bottom, red solid line). The symbols g and * indicate the reflection of insulation layer of the gasket and unknown. The red and green ticks indicate the peak positions of the \ce{H3S} with bcc-structure sulfur and and $\beta$-Po elemental sulfur. The X-ray wavelength was 0.41235 \AA.}
\label{PXRD_integrated}
\end{figure}

\begin{figure}[hbt]
\includegraphics[width=8.6cm]{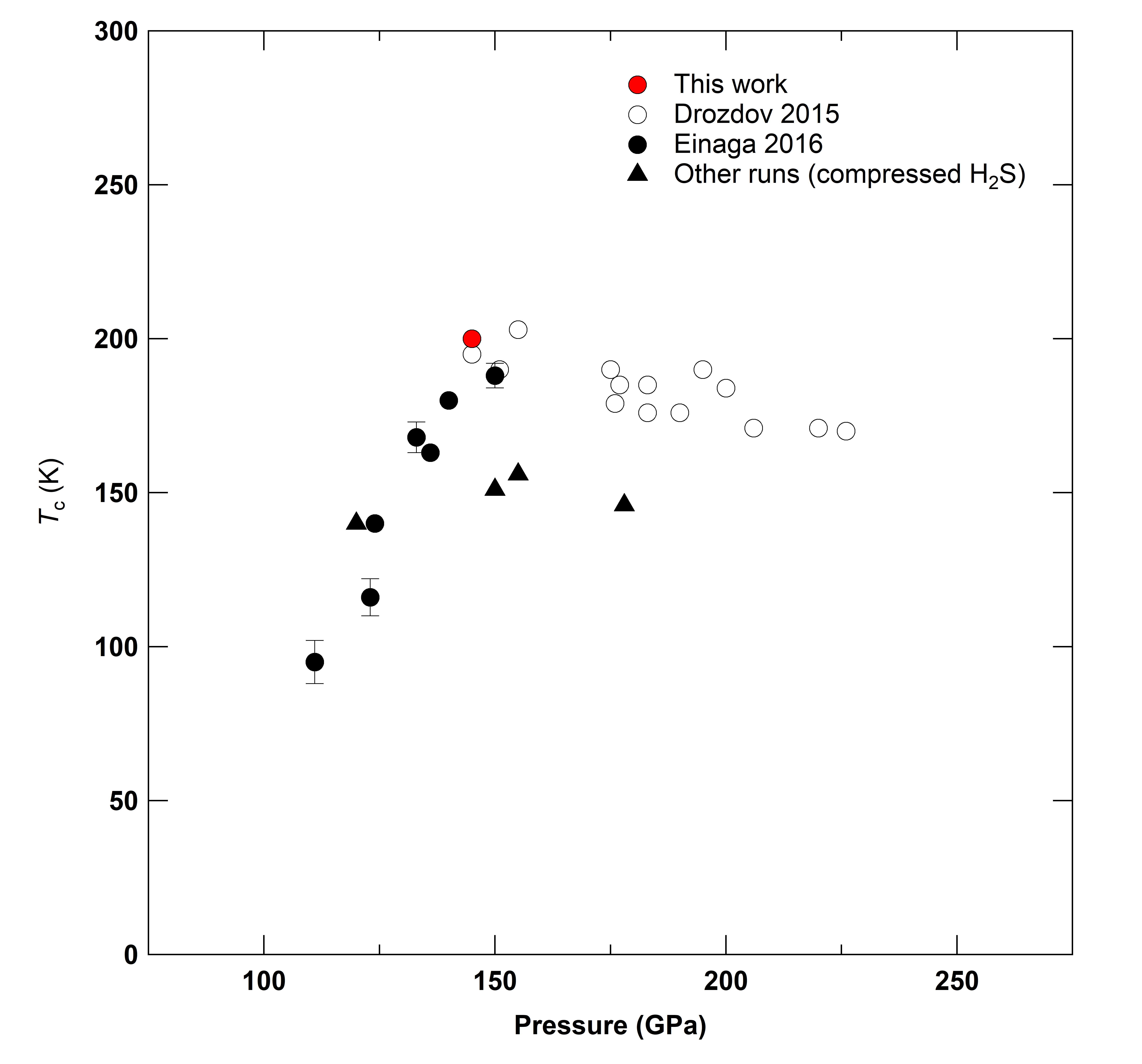}
\centering
\caption{Pressure dependence of \tc\ obtained in previous works with compressed \ce{H2S} by Drozdov \etal\ and our group (open circles from ref \cite{Drozdov2015}, black solid circles from ref \cite{Einaga2016}, black solid triangles from other runs by our group). Red colored circle represents the pressure-temperature point where the sharp step was observed in this study.}
\label{Tc_comparison}
\end{figure}

Superconductivity in sulfur hydride system at high pressure was firstly predicted by Li \etal\ \cite{Li2014} which estimated relatively high \tc\ (\simm80 K) in \ce{H2S}.
However, the experimental results cannot be explained by any stoichiometry \ce{H2S} in terms of \tc.
Soon after Duan \etal\ \cite{Duan2015} published another theoretical work to explore the superconductivity in the S-H system, and found that \ce{H3S} is the most stable stoichiometry under high pressure.
Three structures with the \cccm, \rmm, and \imm\ symmetry are suggested to be realized in experiments and estimated \tc\ for the \rmm\ and \imm\ structure \ce{H3S} are \simm200 K, seems to be consistent to the Drozdov's experimental results \cite{Drozdov2015}.
We have previously performed structural study using synchrotron powder X-ray diffraction (PXRD) and resistance measurements simultaneously in compressed \ce{H2S} \cite{Einaga2016}.
The PXRD patterns obtained in superconducting sample can be explained by mixture of superconductive \ce{H3S} with bcc-structure sulfur sublattice (bcc-like \ce{H3S}) as predicted in Duan's calculation \cite{Duan2015} and elemental sulfur and pressure dependence of \tc\ was in agreement with calculation.
These theoretical and experimental results provided a molecular dissociation scenario where \ce{H2S} decomposes into superconductive \ce{H3S} and sulfur (3\ce{H2S} $\rightarrow$ 2\ce{H3S} + \ce{S}).
Experimentally, \ce{H2S} needs to be compressed at low temperature (below 200 K) to avoid molecular dissociation (phase V of \ce{H2S}) \cite{Fujihisa2004} to reach superconductive \ce{H3S}.
The structural evolution of \ce{H2S} to \ce{H3S} undergoes in sulfur excessive condition, which means possible hydrogen deficiency.
Moreover, PXRD studies revealed that \ce{H3S} synthesized from \ce{H2S} is poorly crystallized \cite{Goncharov2016,Einaga2016} and lower \tc\ can be observed occasionally \cite{Drozdov2015,Einaga2016}.
However, the influence of possible hydrogen deficiency and poor quality of crystalline on superconductivity is still unclear.
Guigue \etal\ and Goncharov \etal\ have perform direct synthesis of \ce{H3S} from elemental sulfur and hydrogen under high pressure and high temperature \cite{Guigue2017,Goncharov2017}.
They loaded mixture of sulfur and hydrogen in diamond anvil cells (DACs) and heated the mixture with infrared laser under high pressure to initiate the chemical reaction of sulfur and hydrogen (2\ce{S} + 3\ce{H2} $\rightarrow$ 2\ce{H3S}).
In their PXRD measurements, it is found that different structures of \ce{H3S} can be synthesized depends on the pressure where the sample is heated - \cccm\ structure for below 140 GPa \cite{Guigue2017} and \imm\ structure for 140 GPa \cite{Goncharov2017}.
Notably, \ce{H3S} with no impurity can be yielded when the reaction undergoes in hydrogen excessive condition, and the quality of the crystalline were much improved compared to compressed \ce{H2S} in previous works \cite{Drozdov2015,Goncharov2016,Einaga2016}.
Due to the major difficulties of hydrogen experiments in DACs, no measurements of superconductivity in directly synthesized \ce{H3S} has been reported. 
In this context, we set the purpose of this study to investigate the superconductivity of directly-synthesized \ce{H3S} by electrical resistance measurements.
We performed experiments in a LH-DAC with 100 $\mu$m culet of diamond anvils for pressure generation.
Rhenium gasket with double-layered insulation layer consists of cubic boron nitride+epoxy and \ce{CaF2} was employed for hydrogen containment.
Diamond anvils were coated by \ce{TiO2} to prevent hydrogen embrittlement.
The thickness of \ce{TiO2} layer was expected to be \simm100 nm.
The sample resistance was measured by AC 2-probe method, and pressure was determined by stress-induced Raman peak of diamond \cite{Akahama2004}.
The sample was prepared as following process; we placed a small piece of sulfur connected to the Au probes.
The DAC was cooled down to \simm11 K in a Gifford-McMahon (GM) refrigerator, and we induced hydrogen gas into the sample chamber at ambient pressure.
As we visibly confirmed the liquefied hydrogen filled the sample chamber, the sample was clamped at 15 K.
The sample was brought back to RT and compressed to 25 GPa to avoid hydrogen leakage.
In Fig.\ref{S+H007_AF_loading}, the microscope photo of the sample is shown.
After loading hydrogen, we continued to compress the sample up to 150 GPa.
During the compression, the electrical resistance of the sample was measured.
The sample resistance largely dropped during the compression, however it remained \simm150 k$\Omega$ at 150 GPa.
This is because of contact resistance caused by the corruption of the probe since sulfur is expected to be a metal above \simm95 GPa \cite{Kometani1997}.
As reported in previous works by Guigue \etal\ \cite{Guigue2017} and Goncharov \etal\ \cite{Goncharov2017}, sulfur hydride was not formed by compression.
The superconducting phase of \ce{H3S} is formed when \ce{S}+\ce{H2} mixture is heated above 140 GPa \cite{Goncharov2017}.
We decided to heat the sample at 150 GPa because the highest \tc\ of the \ce{H3S} was observed at this pressure in previous works with compressed \ce{H2S} \cite{Drozdov2015,Einaga2016}.
We increased the laser power slowly to avoid flashy increase of the sample temperature which may cause diamond failure.
When we applied 12 watt of laser power to the sample, the sample resistance dropped to $\sim$16 $\Omega$.
We measured the temperature during the laser heating by radiation thermometer, and no irradiation was observed when the resistance changed.
Thus the temperature where this transition occurred was T $<$ 1600 K. 
We checked the crystal structure of the sample, and found that superconductive \ce{H3S} with bcc structure sulfur sub-lattice was formed.
The peak positions perfectly fit \imm\ \ce{H3S} and the lattice parameter a = 3.1027(5) \AA\ is in agreement with previous works \cite{Einaga2016,Goncharov2017}.
This drastic change of the sample resistance was likely caused by a partial restoration of the electrical contact between the sample and the Au probe.
The integrated 1-dimensional XRD patterns are shown in Fig.\ref{PXRD_integrated}.
The weak reflections from $\beta$-po sulfur disappeared after laser heating and spotty peaks corresponding to cubic-\ce{H3S} were observed.
The obtained peaks from \ce{H3S} synthesized from elemental sulfur and hydrogen were much sharper compared to \ce{H3S} synthesized from \ce{H2S} yielded in our previous work with compressed \ce{H2S} \cite{Einaga2016}.
Elemental hydrogen still existed after the laser heating, thus we confirmed the \ce{H3S} was synthesized under hydrogen excessive condition.
We carried out electrical resistance measurements in a GM refrigerator.
After laser heating, the pressure was slightly decreased to 146 GPa at RT.
The sample resistance showed metallic behavior and it retained until 200 K.
At 200 K, a sharp step of the sample resistance was observed.
We compared the pressure-temperature data point where the sharp step was observed to the \tc\ obtained in previous works with the compressed \ce{H2S} \cite{Drozdov2015,Einaga2016} (Fig.\ref{Tc_comparison}).
The temperature-pressure point is very close to \tc\ obtained in previous works with compressed \ce{H2S}.
Moreover, no phase transition have been observed in the PXRD measurements \cite{Einaga2016} at low temperature and predicted in theoretical calculations in S-H system at this pressure region.
Therefore we conclude that this sharp step in the sample resistance is corresponding to a superconducting transition.
The onset ($T_{onset}$) and offset ($T_{offset}$) temperature of the superconducting transition was \simm200 K and \simm186 K, respectively. The width of the transition in the resistance was \simm14 K, much narrower than that observed in compressed \ce{H2S} \cite{Drozdov2015,Einaga2016}.
We consider the sharper transition reflects better quality of the crystalline, which means less inhomogeneous of the sample.
We did not observed zero-resistance because we used 2-probe method in this study.
Residual resistance consists of two components, contact resistance and resistance of the Au probes.
The resistivity of the sample estimated from the sample dimension with subtraction of the residual resistance was $\rho$  $\sim1 \times\ 10^{-6}$ $\Omega$$\cdot$m, agreeing the resistivity of  $\rho \sim\ 3 \times\ 10^{-5}-3 \times\ 10^{-7}$ $\Omega$$\cdot$m obtained in the pressure range of the \simm100-200 GPa by Drozdov \etal\ \cite{Drozdov2015}.

In summary, we performed direct synthesis of superconductive \ce{H3S} from elemental sulfur and hydrogen and resistance measurements simultaneously at 150 GPa.
Obtained sample has no excess sulfur and improved crystalline quality as reported in previous works.
A superconducting transition ($T_{onset}$ = 200 K, $T_{offset}$ = 186 K) with a sharp drop of the resistance was observed and $T_{offset}$ = 186 K is the highest $T_{offset}$ in S-H system.

This work was supported by Japan Society for the Promotion of Science Grant-in-Aid for Specially Promoted Research, No. 26000006, JSPS KAKENHI, and performed under proposal No. 2016A1615, 2017B1711 and 2018A0149 of the SPring-8.

\begin{figure}[hbt]
\includegraphics[width=8.6cm]{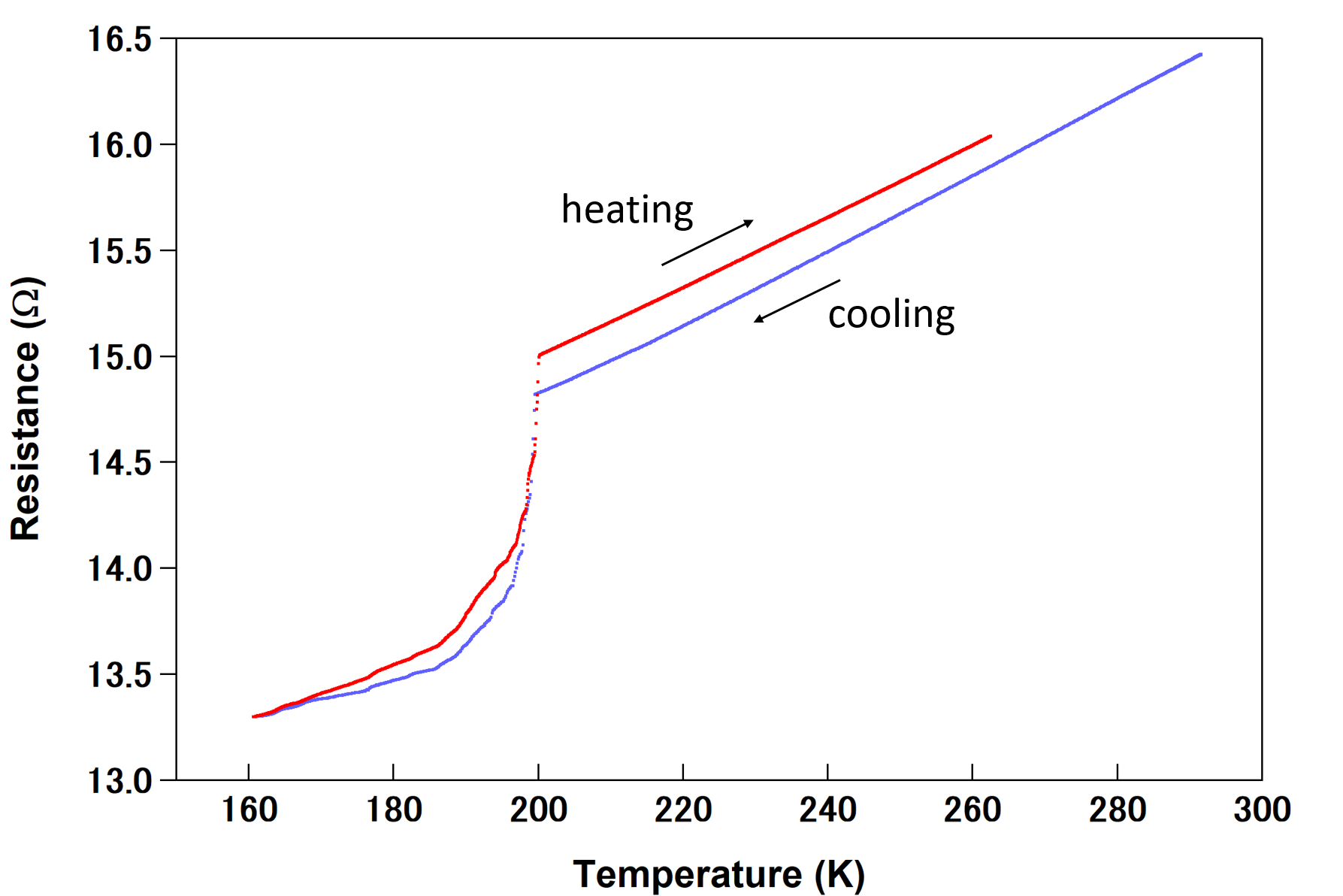}
\centering
\caption{Temperature dependence of the sample resistance after laser heating (146 GPa). Red and blue lines represent the sample resistance in the heating process and cooling process, respectively.}
\label{R_C_T}
\end{figure}
\end{document}